\documentclass[journal,comsoc]{IEEEtran}
\usepackage{cite}

\usepackage{ifpdf}
\ifpdf
	\usepackage[pdftex]{graphicx}
\else
	\usepackage[dvips]{graphicx}
\fi

\usepackage{amsmath}
\usepackage{url}
\usepackage{gb4e}

\usepackage{algorithm}
\usepackage{algorithmic}

\begin{document}
\title{Aesop Fable for Network Loops}

\author{Marc Mosko, Glenn Scott, and Dave Oran
\thanks{M. Mosko and G. Scott are with Palo Alto Research Center (PARC)}
}

\markboth{}%
{Mosko \MakeLowercase{\textit{et al.}}: Aesop Fable of Network Loops}

\maketitle

\begin{abstract}
Detecting loops in data networks usually involves counting down a hop limit or caching
data at each hop to detect a cycle.  Using a hop limit means that the origin of a packet must know
the maximum distance a packet could travel without loops.  It also means a loop is not detected
until it travels that maximum distance, even if that is many loops.  Caching a packet signature at each
hop, such as a hash or nonce, could require large amounts of memory at every hop because that
cached information must persist for as long as a loop could forward packets.  This paper presents
a new distributed loop detection mechanism based on a Tortoise and Hare algorithm that can quickly
detect loops without caching per-packet data at each hop with a modest amount of additional state
in each packet.
\end{abstract}

\IEEEpeerreviewmaketitle

\section{Introduction} 

Detecting loops in data networks usually involves counting down a hop limit or caching
data at each hop to detect a cycle.  Using a hop limit means that the origin of a packet must know
the maximum distance a packet could travel without loops.  It also means a loop is not detected
until it travels that maximum distance, even if that is many loops.  Caching a packet signature at each
hop, such as a hash or nonce, could require large amounts of memory at every hop because that
cached information must persist for as long as a loop could forward packets.  This paper presents
a new distributed loop detection mechanism based on a Tortoise and Hare algorithm that can quickly
detect loops without caching per-packet data at each hop with a modest amount of additional state
in each packet.

One class of algorithms for detecting loops in a series are based on the tortoise and hare 
construction, attributed to Floyd by Knuth~\cite{knuth}.  The core idea is that the tortoise moves
one-by-one through a series, denoted as $x_i$, and the hare moves twice as fast, denoted as $x_{2i}$.
If there is a cycle, eventually $x_{2i} = x_i$, where hop $2i$, modulo the cycle length, is the same item
as $i$.  At each step of the algorithm, $tortoise \leftarrow \operatorname{next}(tortoise)$ and $hare \leftarrow \operatorname{next}( \operatorname{next} (hare))$.

Brent's Algorithm~\cite{brent} finds cycles like Floyd's algorithm, however it requires less memory
than Floyd's algorithm to compute the ``next'' operation.  Brent's algorithm compares $x_{2^i-1}$ with
all $x_j$ in the open interval $j = [2^i, 2^{i+1})$.  Therefore, the same $x_{2^i-1}$ is used at each
iteration until there's a single update.  $tortoise \leftarrow x_{2^i-1}$, $hare \leftarrow \operatorname{next}(hare)$,
until the next power of two when $tortoise \leftarrow hare$.  This formation keeps the state update
using ``next'' proceeding in-step with the algorithm iteration, so a packet does not need to carry any memory
with it beyond $hare$.

Section~\ref{sec:brent} reviews Brent's centralized algorithm.  Section~\ref{sec:distributed} presents
our distributed version of the algorithm for use in a packet network.

\section{Brent's Algorithm}\label{sec:brent}

Algorithm~\ref{alg:brent} shows Brent's algorithm for the case where $\mathcal{Q}=2$ and $\mu=0$, which
he calls the common case, based on~\cite{brent}.  We have used more descriptive names than $x, y, k, r$.
Mapping Brent's variables to more descriptive names, we use $x \rightarrow hare$, $y \rightarrow tortoise$, $r \rightarrow power$, and $k \rightarrow hops$.

In this formulation, the algorithm tracks four variables: $tortoise$, $hare$, $power$, and $hops$.  For a centralized
algorithm, tracking this information is not a burden, but in a distributed algorithm we wish to minimize the 
message sizes to reduce overhead.  In our distributed version of the algorithm, in Section~\ref{sec:distributed},
we reduce the necessary message size to only $tortoise$ and $hops$.

\begin{algorithm}[t]
\caption{Brent's Algorithm (for base 2)}
\label{alg:brent}
\begin{algorithmic}[1]
\STATE $hare \gets x_0$
\STATE $power \gets 1$
\STATE $hops \gets 0$
\REPEAT
	\STATE $tortoise \gets hare$
	\STATE $power \gets power * 2$
	\REPEAT
		\STATE $hops \gets hops + 1$
		\STATE $hare \gets \operatorname{next}(hare)$
	\UNTIL $tortoise = hare$ \OR $hops \ge power$ \OR $hare = \emptyset$
\UNTIL $tortoise = hare$ \OR $hare = \emptyset$
\IF{$tortoise = hare$}
	\RETURN true
\ENDIF
\RETURN false
\end{algorithmic}
\end{algorithm}

\section{Distributed Brent's Algorithm}\label{sec:distributed}
Our use of Brent's algorithm is shown in Alg~\ref{alg:init} and Alg.~\ref{alg:receive}.  When a node
creates a new packet, it initializes the $tortoise$ variable to its node identifier (nodeid).
The node identifier could be the hash a a system's public key or an administratively assigned
unique identifier.
When a node receives a packet, it first increments the hop count.  If the packet's $tortoise$
is equal to the current node's $nodeid$, then there is a loop (the algorithm returns true).
Otherwise, if the hop count is equal to the $power$, then the packet has traveled a power of 2
hops, so the algorithm updates the $tortoise$ value.

Based on the hop count, one can detect if the packet has gone a power of 2 distance using the
equation \ref{powerof2}, a commonly known bit manipulation, where the symbol $\&$ means
bit-wise AND.

\begin{equation}\label{powerof2}
\operatorname{hops \; is \; power \; of \;2} \Longleftrightarrow ( hops \; \& \; (hops-1) ) = 0
\end{equation}

\begin{algorithm}[t]
\caption{Initialize Packet}
\label{alg:init}
\begin{algorithmic}[1]
\STATE $packet.tortoise \gets nodeid$
\STATE $packet.hops \gets 0$
\end{algorithmic}
\end{algorithm}

\begin{algorithm}[t]
\caption{Receive Packet}
\label{alg:receive}
\begin{algorithmic}[1]
\STATE $packet.hops \gets packet.hops + 1$
\IF{$packet.tortoise = nodeid$}
	\RETURN true
\ELSIF{$packet.hops$ is power of 2}
	\STATE $packet.tortoise \gets nodeid$
\ENDIF
\RETURN false
\end{algorithmic}
\end{algorithm}

\begin{figure}
\centering
\includegraphics[width=\linewidth, clip, page=1, trim=0.4in 1.0in 1.0in 1.0in]{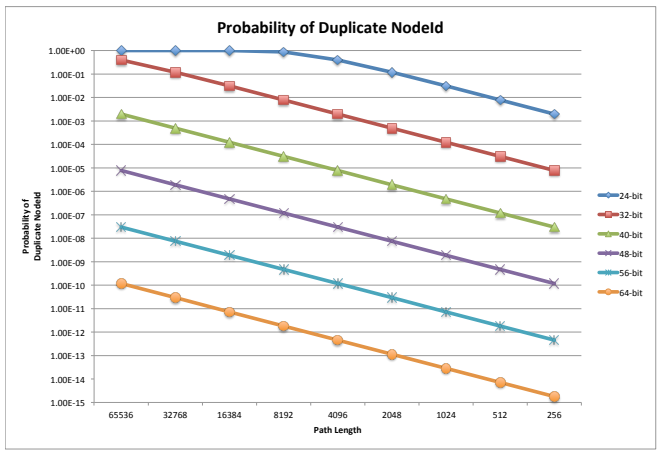}
\caption{Probability of duplicates on a path}
\label{fig:object}
\end{figure}

The size of $tortoise$ must be large enough that there is a very low probability of duplicates along
a given path.   Either the size must be so large that there's a vanishingly small probability that 
there are no duplicates or we should ensure that a retransmission will not have the same failure mode.
Fig.~\ref{fig:object} uses the Birthday Paradox to calculate the probability that two or more routers
on a path of Path Length have the same NodeId assuming the IDs are picked randomly.
For example, for a 32-bit nodeid, there's about a 1\% chance of collision for a path of 8192 hops.
For paths of practical maximum lengths, such as at most 256 to 512 hops, node ids of 48 to 64 bits should
be sufficient.

If we wish to make sure that a retransmission breaks a path with duplicate node IDs, we can need to use a virtual node ID.  
One could calculate the virtual node ID as 
$$id = \operatorname{hash}(trueid, \operatorname{hash}(packet))$$, 
where the $trueid$
is a large number, such as the SHA-256 hash of a system's public key, and the hash of the packet includes
a short nonce per retransmission and does not include the loop-prevention fields (tortoise, power, hops).
This method also better preservers anonymity because the $tortoise$ identity is scrambled, potentially with
a cryptographic-grade hash.


\begin{thebibliography}{1}
\providecommand{\url}[1]{#1}
\csname url@samestyle\endcsname
\providecommand{\newblock}{\relax}
\providecommand{\bibinfo}[2]{#2}
\providecommand{\BIBentrySTDinterwordspacing}{\spaceskip=0pt\relax}
\providecommand{\BIBentryALTinterwordstretchfactor}{4}
\providecommand{\BIBentryALTinterwordspacing}{\spaceskip=\fontdimen2\font plus
\BIBentryALTinterwordstretchfactor\fontdimen3\font minus
  \fontdimen4\font\relax}
\providecommand{\BIBforeignlanguage}[2]{{%
\expandafter\ifx\csname l@#1\endcsname\relax
\typeout{** WARNING: IEEEtran.bst: No hyphenation pattern has been}%
\typeout{** loaded for the language `#1'. Using the pattern for}%
\typeout{** the default language instead.}%
\else
\language=\csname l@#1\endcsname
\fi
#2}}
\providecommand{\BIBdecl}{\relax}
\BIBdecl

\bibitem{knuth}
D.~Knuth, \emph{The Art of Computer Programming}.\hskip 1em plus 0.5em minus
  0.4em\relax Addison-Wesley, 1969, vol.~2.

\bibitem{brent}
\BIBentryALTinterwordspacing
R.~Brent, ``\BIBforeignlanguage{English}{An improved monte carlo factorization
  algorithm},'' \emph{\BIBforeignlanguage{English}{BIT Numerical Mathematics}},
  vol.~20, no.~2, pp. 176--184, 1980. [Online]. Available:
  \url{http://dx.doi.org/10.1007/BF01933190}
\BIBentrySTDinterwordspacing

\end{thebibliography}


\end{document}